\documentclass[twoside]{article}
\usepackage{fleqn,espcrc2,epsfig}

\usepackage{graphicx}
\usepackage[figuresright]{rotating}

\newcommand{\AmS}{{\protect\the\textfont2
  A\kern-.1667em\lower.5ex\hbox{M}\kern-.125emS}}

\title{Quantum Protectorates in the Cuprate Superconductors}

\author{David Pines\address{ 
Institute for Complex Adaptive Matter,
University of California Office of the President, \\
LANSCE Division,
Los Alamos National Laboratory, and
Science and Technology Center for Superconductivity, University of
Illinois
}}
\begin{document}
\setlength{\topmargin}{0cm}
\begin{abstract}
  Following the identification of the pairing state, the major
  challenge in understanding the cuprate superconductors has been
  determining the evolution with doping and temperature of their
  anomalous normal state behavior. Key to this understanding is the
  experimentally determined magnetic phase diagram for the cuprates,
  which provides information on the protected magnetic properties of
  the normal state, generic behavior that is reliably the same one
  system to the next, regardless of details. I discuss the constraints
  this places on candidate quantum protectorates, and the status of
  microscopic model calculations for a protectorate consistent with
  these constraints, the nearly antiferromagnetic Fermi liquid.
\end{abstract}
\maketitle

\section{Introduction}

In seeking a theoretical understanding of the cuprate superconductors,
a key question from the outset has been the relevance of attempts to
deduce their behavior from first principles.  As Laughlin and I have
recently argued \cite{toe}, ab-initio computations have failed completely to
explain their phenomenology; indeed it would appear that not only has
deduction from microscopics not been able to explain the wealth of
crossover behavior found in the underdoped cuprates, but that as a
matter of principle it probably cannot explain it, much less calculate
the high transition temperatures found at optimal doping. We concluded
that a more appropriate starting point would be to focus on the
results of experiments on the low energy properties of the novel
states of matter found in the cuprates in the hope of identifying the
corresponding quantum protectorates--stable states of matter whose
generic low energy properties, insensitive to microscopics, are
determined by a higher organizing principle and nothing else.  To the
extent one has correctly identified the quantum protectorate, one
would then hope that microscopic protectorate-based toy model
calculations might be relevant for understanding experiment. From this
perspective, in conventional superconductors one observes a transition
from a Landau Fermi liquid protectorate to the BCS s-wave
protectorate; each protectorate can be characterized by a small number
of parameters, which can be determined experimentally, but which are,
in general, impossible to calcuate from first principles.

In the cuprate superconductors both the normal state
and the superconducting state protectorates differ dramatically from
their conventional superconductor counterparts.  There is now a
consensus that the superconducting protectorate is a BCS d-wave
protectorate, but no consensus has been reached on the nature of the
novel normal state phases. Candidate protectorates that have been
proposed include
Luttinger liquids, nearly antiferromagnetic Fermi liquids, nearly
charge-ordered Fermi liquids, mesoscopically ordered phases
(stripes), and quantum critical behavior.In this talk I  will present
the case for one of these, the nearly antiferromagnetic Fermi liquid
(NAFL) protectorate.

\section{Experimental evidence for the
NAFL protectorate and
dynamical scaling}

For over a decade it has been known from NMR measurements that a
single spin component is responsible for the planar $^{63}$Cu and
$^{17}$O spin-lattice relaxation rates and Knight shifts, and that a
quantitative account of these experiments, as well as the more recent
experiments on the $^{63}$Cu spin-echo decay time, may be obtained
with the generic low energy dynamic magnetic susceptibility
appropriate for a commensurate almost antiferromagnetic protectorate
\cite{mmp},
\begin{equation}
  \label{eq:mmp}
  \chi({\bf q},\omega) = \frac{\alpha \xi^2}{1+\xi^2({\bf q-Q})^2 -
    i\omega_{sf}/\omega}
\end{equation}
where  $Q =(\pi,\pi)$ is the commensurate wavevector, $\xi$
is the antiferromagnetic
correlation length, $\omega_{sf}$ is the frequency of the relaxational
mode, and
$\alpha$ is the scale factor
that relates the static commensurate susceptibility, $\chi_Q$,
to the square
of the  correlation  length, $\chi_Q=\alpha \xi^2$.
 For the optimally-doped 1-2-3 system,near $T_c$,
$\xi\sim$ twice the lattice constant, a, $\alpha$ is $\sim
15$ states/eV, so $\chi_Q$ is some 60 states/ev,
large indeed compared to the expected Landau Fermi liquid
value of $\sim 1$ state/ev; the corresponding value of
$\omega_{sf}$
is some
15 meV, small compared to the Landau Fermi liquid
value of $\sim 1$ eV.
Still larger values of $\chi_Q$ and $\xi$, and smaller values of
$\omega_{sf}$ are encountered as one goes to the underdoped materials.

As sample quality and the range of experiments improved, it became
possible for Barzykin and Pines \cite{bp} to construct the generic
magnetic phase diagram for 1-2-3, Bi-based, and Hg-based materials
shown in Fig.~1 and subsequently confirmed in the NMR experiments of
Curro et al \cite{charlie} on YBa$_2$Cu$_4$O$_8$
and Aeppli \emph{et al } \cite{aeppli}
on nearly optimally doped 2-1-4.  For magnetically underdoped
materials,those that exhibit a maximum in the uniform
temperature-dependent susceptibility at some temperature $T_{cr}$,
experiment shows that at $T_{cr}$, $\xi$ is $\sim2$,and the system
crosses over from mean field $z=2$ behavior to a $ z=1$ dynamical
scaling regime. The corresponding weak pseudogap behavior persists
until a second crossover in the normal state takes place at $T_*$, the
temperature at which $^{63}$Cu $T_1$T is minimum; in this strong
pseudogap regime, one no longer has $z=1$ scaling, while ARPES
experiments show that an energy gap develops for the quasiparticles
near $(\pi,0)$. In the 2-1-4 system, there is little or no evidence
from spin-lattice relaxation rates for a crossover to strong pseudogap
behavior. Magnetically overdoped materials do not exhibit any
crossover behavior in the normal state; the af correlations never
become strong enough to bring about the weak pseudogap behavior
associated with nascent spin density wave formation and $z=1$
dynamical scaling, and one finds a direct transition from normal state
mean field behavior to superconductifvity.
\begin{figure}[t]
    \epsfig{file=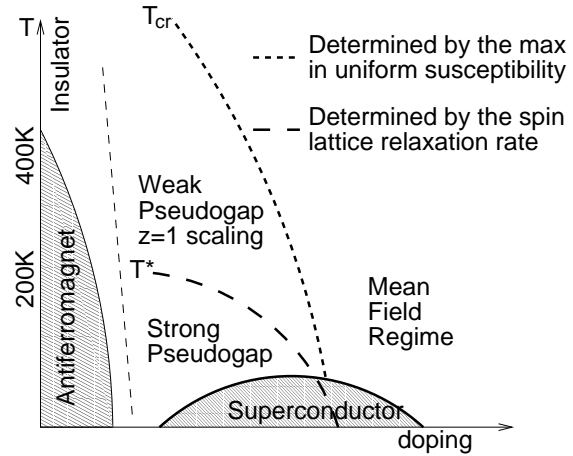,angle=0,width=7.50cm}
    \caption{Generic phase diagram of high temperature superconducting
      cuprates. The true thermodynamic phases (antiferromagnetic at
      low doping and superconducting at higher doping are depicted
      by the shaded regions. The remaining lines correspond to
      crossovers, visible in a variety of experiments.}
\end{figure}

\section{Theoretical support for the NAFL protectorate}

Since transport and specific heat measurements suggested that planar
quasiparticles of some kind were determining the properties of the
normal state, our theoretical group in Urbana was led to ask what
might be the properties of a Fermi liquid which exhibits nearly
antiferromagnetic behavior. We carried out microscopic
calculations for a relevant toy model-- quasiparticles, whose spectra
were characterized by nearest neighbor and next-nearest neighbor
hopping terms, interacting through spin-fluctuation exchange, with an
effective interaction that is proportional to the dynamic magnetic
susceptibility and hence reflects the near approach to
antiferromagnetism required by the NMR experiments,
\begin{equation}
  \label{eq:veff}
  V_{eff} = g^2 \chi({\bf q},\omega)
\end{equation}
where $g$ is an arbitrary coupling constant.

Weak coupling numerical calculations \cite{mbp} of the consequences
of this experiment-based highly anisotropic quasiparticle interaction
(note that only quasiparticles in the vicinity of the "hot spots",
regions of the Fermi surface separated by {\bf Q},
feel the full effects of the interaction) showed that the
resulting resistivity would be roughly linear in
temperature, and that for modest values of the coupling constant one
could easily get a transition at high temperatures to a
superconducting state with $d_{x^2-y^2}$ pairing. Subsequent strong coupling
(Eliashberg) calculations \cite{mp} showed that for a coupling constant
which yielded quantitative agreement with experiment for the magnitude
and temperature dependence of the resistivity of optimally-doped
1-2-3, a transition to the $d_{x^2-y^2}$ state took place at $\sim 100$K.
Armed
with this "proof of concept" for the candidate NAFL protectorate,
Monthoux and I predicted that over time experimentalists would find
our calculated pairing state, which indeed proved to be the case
within the next few years. (Had they not done so we were prepared to
abandon our NAFL approach.)  A second prediction was highly
anisotropic quasiparticle behavior as one moves around the Fermi
surface; the calculated frequency and temperature dependent
self-energy of the hot quasiparticles located near $(\pi,\pi)$
was highly
anomalous, with an imaginary part which was proportional to the
maximum of $\omega$ or $T$, while that of the cold quasiparticles, those
near the diagonals from $(0,0)$ to $(\pi,\pi)$ that feel  little of the NAFL
interaction, was Landau-Fermi-liquid like. a prediction that was
borne out by subsequent ARPES experiments on overdoped materials.

This highly anisotropic quasiparticle behavior of hot and cold
quasiparticles was shown by Stojkovic  \cite{sp}
to provide a natural
explanation for the measured anomalous Hall transport and optical
behavior of the overdoped materials. Using a Boltzmann equation
approach, Stojkovic found that the ctn of the Hall angle
was determined almost entirely by the cold quasiparticles, and so
would necessarily exhibit $T^2$ behavior for a wide range of dopings,
while both hot and cold quasiparticles contribute to the conductivity
in such a way as to yield the familiar linear in T behavior of the
resistivity. Stojkovic obtained quantitative agreement with
experiments on the conductivity, Hall conductivity, optical
conductivity, magnetotransport, and thermoelectric behavior of the
overdoped and underdoped systems. Subsequent detailed calculations by
Monthoux \cite{twoloop}
have shown that as long as one is in the mean-field regime,
vertex corrections to
the Eliashberg calculations are small, and  do not bring about an
appreciable change in the mean field calculations of transport
properties,

The success of these calculations makes a very strong case for the
proposition that in magnetically overdoped systems the normal state is
an NAFL protectorate with a dynamic magnetic susceptibilty which
exhibits $z=2$ dynamical scaling behavior. Matters are  otherwise
for magnetically underdoped systems, where (see Fig.\ 1) weak
pseudogap behavior is found at temperatures below $T_{cr}$,
where the dynamic magnetic
susceptibility exhibits $z=1$ dynamical scaling behavior,
while at still lower temperatures one
finds a second crossover at $T_*$ to strong pseudogap behavior. Still it
 proved possible to extend the NAFL toy model calculations to cover
the weak pseudogap regime ($T_*<T<T_{cr}$) where Schmalian
\emph{et al} \cite{cps} have shown that in the classical limit of
temperatures large
compared to the spin fluctuation energy,$\omega_{sf}$, one can sum all the
relevant diagrams in a perturbation-theoretic treatment of the
interaction, Eq.\ (2); they find, in agreement with experiment, a
substantial transfer of spectral weight from low to high frequencies
for the hot quasiparticles, a cross-over from $z=2$ to $z=1$ scaling
behavior at $\xi\sim 2a$, and a uniform susceptibility that decreases as
the temperature decreases, and show that all these phenomena are
associated with the nascent spin density wave formation anticipated
for longer AF correlation lengths. They found the
renormalized spin fluctuation-quasiparticle coupling is enhanced
at low frequencies by vertex corrections, contrary to an argument
presented by Schrieffer, but that at high frequencies it is reduced,
in agreement with Schrieffer. Finally, Schmalian \cite{rg} has developed an
RG approach to the NAFL which takes into account the damping of spin
waves by quasiparticle-quasihole pairs. He finds the crossovers from
quantum disordered to $z=1$
quantum critical to mean field behavior that are seen experimentally, while
his calculated  temperature dependence of the AF
correlation length  agrees with that obtained  by Aeppli \emph{et al}
\cite{aeppli} in their INS experiments on near-optimally-doped 2-1-4.

\section{Open questions}

Despite the evident success of theoretical calculations based on the
NAFL protectorate, there remain a number of open issues and questions.
Foremost, what is the protectorate in the
strong pseudogap regime?  Experiment tells us that the leading edge
gap measured for the hot quasiparticles in this regime develops
rapidly as the temperature is lowered below $T_*$, while its doping
dependence is anomalous, increasing  with decreasing doping.
This gap thus has a markedly different doping dependence than that
found for the cold quasiparticles; the latter tracks $T_c$, which
decreases as the doping is reduced. Until one understands this
protectorate no model calculation of the superconducting transition
temperature for the underdoped cuprates is possible. A second
question, for the 2-1-4 system, is the origin of the incipient, and
typically dynamic, mesoscopic ordering of spin and charge inferred
from the appearance of incommensurate peaks in the INS measurements of
spin fluctuations. Is this ordering  related to the failure to find any
magnetic
evidence for strong pseudogap behavior in these materials? Clearly for
this system one needs to sort out the consequences, including quantum
critical behavior, of the competition between antiferromagnetism,
superconductivity, and stripe formation. Why, for similar doping
levels, do the 2-1-4 materials have substantially lower
superconducting transition temperatures than the 1-2-3, Bi-based, or
Hg-based systems? Does their greater tendency toward mesoscopic
ordering inhibit superconductity? But why do they possess this greater
tendency? And why do INS and NMR experiments exhibit no sign of strong
pseudogap behavior (where is the $T_*$?).

In summary, it can reasonably be argued that we have been able to
identify the protectorates associated with two of the three normal
state phases found in magnetically underdoped superconductors--the $z=2$
NAFL
mean field protectorate found above $T_{cr}$ and the $z=1$ NAFL
protectorate found
between $T_{cr}$ and $T_*$, and to demonstrate that for these protectorates
there is no spin-charge separation. But the issue of the strong
pseudogap protectorate remains open, and until it is solved, and the
above questions answered, we will be far from possessing  a complete
understanding of these remarkable systems.

\end{document}